\documentstyle[amsfonts,aps]{revtex}

\begin{document}
\author{M. Castagnino}
\address{Departamento de F\'{i}sica, Universidad de Buenos Aires\\
Casilla de Correos 67, Sucursal 28, 1428\\
Buenos Aires, Argentina}
\author{M. Gadella}
\address{Departamento de F\'{\i }sica Te\'{o}rica, Universidad de Valladolid\\
c. Real de Burgos, s.n.,47011\\
Valladolid, Spain.}
\title{The role of self-induced decoherence in the problem of the classical limit
of quantum mechanics}
\maketitle

\begin{abstract}
Our account of the problem of the classical limit of quantum mechanics
involves two elements. The first one is self-induced decoherence, conceived
as a process that depends on the own dynamics of a closed quantum system
governed by a Hamiltonian with continuous spectrum; the study of decoherence
is addressed by means of a formalism used to give meaning to the van Hove
states with diagonal singularities. The second element is macroscopicity
represented by the limit $\hbar \rightarrow 0$: when the macroscopic limit
is applied to the Wigner transformation of the diagonal state resulting from
decoherence, the description of the quantum system becomes equivalent to the
description of an ensemble of classical trajectories on phase space weighted
by their corresponding probabilities.
\end{abstract}

\section{Introduction}

The problem of the classical limit of quantum mechanics has been a point of
debate since the birth of the theory. Although this problem is usually
addressed in the context of measurement, it can be analyzed from a more
general point of view, in terms of {\it how the classical world arises from
an underlying quantum reality}, independently of whether there is a
measurement involved or not. Of course, the problem of the classical limit
relies on the assumption that, if quantum mechanics is correct, then its
results must reproduce the results of classical mechanics in the appropriate
limit.

In the old days of the theory, Heisenberg and Bohr among others conceived
the classical limit of quantum mechanics by analogy with the classical limit
of special relativity: $\hbar \rightarrow 0$ in quantum mechanics should
play the same role as $\beta \rightarrow 0$ in special relativity. This
assumption was considered by Einstein as an oversimplification since, while
relativity and classical mechanics have the same deterministic structure,
quantum mechanics has a probabilistic structure. Nevertheless, since those
days it has been usually claimed that classical mechanics can be recovered
as a limiting case of quantum mechanics when $\hbar \rightarrow 0$. This
assumption led to correct results when the classical limit was conceived in
the following way:

\begin{equation}
{\rm QM}\left\{ 
\begin{array}{l}
\text{ }\overleftarrow{\text{ \qquad \qquad }quantization\qquad \qquad } \\ 
\text{ }\overrightarrow{\quad \quad \text{{\it \ classical limit}}\equiv
\hbar \rightarrow 0\text{{\it \quad }}}
\end{array}
\right\} {\rm CM}  \label{S.1}
\end{equation}
where QM and CM stand for quantum mechanics and classical mechanics
respectively. In this schema, the first step is to quantize a classical
system, e.g., by means of the Weyl transformation, in order to obtain the
corresponding quantum system (at present, quantization is also called
''deformation''). Then, the original classical system is recovered by
applying the inverse Weyl transformation, i.e., the Wigner transformation%
\footnote{%
Historically, Weyl proposed his transformation as a quantization method.
Later and independently, Wigner proposed a transformation that mapped
quantum states into classical density functions. Finally, Moyal proved that
the Wigner transformation was equivalent to the inverse Weyl transformation.}
to the quantum system previously obtained and by taking the limit $\hbar
\rightarrow 0$. It is quite clear that this method is completely circular to
the extent that it only recovers the classical system originally proposed.

When the theoretical structure of quantum mechanics finally lost its
classical origin, the problem of the classical limit acquired a new
formulation that became the traditional one:

\begin{equation}
{\rm QM}\left\{ \overrightarrow{\quad \quad \quad \text{{\it \ classical
limit}}\equiv \hbar \rightarrow 0\text{{\it \quad \quad \quad }}}\right\} 
{\rm CM}  \label{S.2}
\end{equation}
This means that, no matter how the original quantum system was described, a
classical system should be obtained via the Wigner transformation when $%
\hbar \rightarrow 0$. However, this way of conceiving the problem of the
classical limit leads, at least, to three problems:

\begin{enumerate}
\item  \smallskip In general, the Wigner state function $\rho (\phi )$
(where $\phi =(q,p)$ is a point in phase space) is not non-negatively defined%
\footnote{%
A function $\rho (\phi )$ is non-negative if and only if $\rho (\phi )\geq 0$%
.}; as a result, it cannot be interpreted as a probability distribution.

\item  \smallskip Only Hamiltonians of degree $\leq 2$ in $p$ and $q$ yield
to Hamiltonian fluxes that maintain the deformation invariant (or covariant)
(see \cite{DS}, \cite{BFFLS}). In fact, only in these cases $\rho _{1}\rho
_{2}(t)=\rho _{1}(t)\rho _{2}(t)$ after performing the Wigner transformation.

\item  In some cases, factors of the form $\hbar ^{-1}$ may appear in the
Wigner state function due to the features of the Wigner transformation (see 
\cite{DS}, \cite{Ballentine}). In these cases, the limit $\hbar \rightarrow 0
$ of the Wigner function is singular.
\end{enumerate}

In this paper we will follow a well known trend in contemporary physics,
according to which the classical limit must not be conceived as a mere
consequence of a limiting procedure, but as a result of a physical process.
From this perspective, the explanation of the emergence of the classical
world from the underlying quantum realm involves two steps: the first one
consists in explaining the physical phenomenon of decoherence, and the
second one consists in taking the macroscopic limit $\hbar \rightarrow 0$.
However, we will move away from the mainstream position with respect to the
explanation of decoherence: the aim of this paper is to obtain the classical
limit of quantum mechanics on the basis of the {\it self-induced approach to
decoherence}, such as it was presented in paper \cite{Deco} and discussed in
depth in paper \cite{CL}. In contrast to the traditional einselection
approach \cite{JP}, from the self-induced perspective decoherence does not
require the openness of the system and its interaction with the environment:
a single closed system can decohere when it has continuous spectrum. We will
show that, in this new scenario, the classical limit is described by the
following diagram:

\begin{equation}
{\rm QM}\left\{ 
\begin{array}{c}
\overrightarrow{\quad \quad \quad \text{{\it \ decoherence\quad \quad \quad }%
}}{\rm Boolean\ QM}\text{ }\overrightarrow{\quad \text{{\it \ \quad
macroscopicity}}\equiv \hbar \rightarrow 0\text{{\it \quad \quad \quad }}}
\\ 
\overrightarrow{\qquad \qquad \qquad \qquad \qquad \qquad \quad \qquad \text{%
{\it classical limit}}\qquad \qquad \qquad \qquad \qquad \qquad \qquad }
\end{array}
\right\} {\rm CSM}  \label{S.3}
\end{equation}
Self-induced decoherence transforms quantum mechanics into a Boolean quantum
mechanics where the interference terms that preclude classicality have
vanished. Macroscopicity, expressed by the limit $\hbar \rightarrow 0$%
\footnote{%
It is quite clear that it is not possible to set the value of $\hbar $ equal
to $0$, since it is not a dimensionless parameter but an universal constant.
This means that, strictly speaking, the macroscopic limit is $\hbar
/S\rightarrow 0$, where $S$ is the characteristic action of the system: this
is a factual limit which represents realistic situations where $S>>\hbar $.}%
, turns Boolean quantum mechanics into classical statistical mechanics (CSM)
in phase space. According to this view, the classical limit of quantum
mechanics is not classical mechanics but {\it classical statistical mechanics%
}, and it requires two physical conditions: {\it decoherence} and {\it %
macroscopicity}. In other words, in order to behave classically a quantum
system must have decohered and must be macroscopic enough: each one of these
conditions alone is necessary but not sufficient for its classical behavior.
Furthermore, we will show how and under what conditions this explanation
overcomes the three problems that arise from the traditional way of
conceiving the classical limit.

This paper is organized as follows. In Section II we present the formalism
for observables and states, necessary for developing our program. Section
III is devoted to explain the self-induced approach to decoherence:
decoherence in energy and in the remaining variables are considered. In
Section IV we study the operation known as Wigner transformation and its
application to observables and states. In Section V we show how the
classical limit leads to classical statistical mechanics when the
macroscopic limit is applied to the Wigner transformation of the quantum
state resulting from decoherence. In Section VI we discuss the physical
meaning of the results just obtained, arguing that classicality must be
understood as an emergent property that objectively arises from an
underlying quantum mechanical realm. Finally, in Section VII we draw our
conclusions. Two appendices complete the paper.

\section{Formalism for observables and states}

The formalism for observables and states used in the present paper is
inspired by the formalism introduced by Antoniou {\it et al.} (\cite{AS3}, 
\cite{ASLT}) which, in turn, is based on the works of van Hove \cite{vh}. In
several papers (\cite{Pra}, \cite{Pre}, \cite{Petrus}, \cite{CG}, \cite
{Gamov}, \cite{Rolo}) we applied different versions of this formalism to the
study of the properties of quantum systems with continuous spectrum. In
particular, the formalism was used in paper \cite{Deco} for explaining
decoherence. In order to simplify the notation, here we will study a
simplified model where all the observables have continuous spectrum (cases
where all observables except $H$ have discrete spectrum will be considered
in the footnotes): this will allow us to improve the mathematical basis of
our approach without a proliferation of indices that would not introduce
conceptual advantages.

\subsection{Quantum operator algebra}

Let us consider a system with a complete set of commuting observables (CSCO) 
$\{H,O_{1},...,O_{N}\}$ where $H$ has a continuous spectrum $0\leq \omega
<\infty $ and, for the sake of simplicity, the $O_{i}$, $i=1,2,\dots ,N$,
have also continuous spectra\footnote{%
The continuous spectrum is relevant for the classical limit since, in the
limit $\hbar \rightarrow 0$ (precisely, the high quantum number limit), many
discrete spectra become continuous. Spectra with continuous and discrete
parts are studied in papers \cite{Deco} and \cite{Rolo}.}. We will assume
that the observables $H,O{_{1}},...,O_{N}$ are Weyl observables, i.e., that
they come from the Weyl transformation of classical observables. In order to
simplify the notation we will use $\{H,O\}$ to denote the CSCO $%
\{H,O_{1},...,O_{N}\}$. The generalized eigenbasis of $\{H,O\}$ is $%
\{|\omega ,o\rangle \}$, where $\omega $ and $o$ satisfy:

\begin{equation}
H\,|\omega ,o\rangle =\omega \,|\omega ,o\rangle \quad \quad {\rm and\quad }%
\quad O\,|\omega ,o\rangle =o\,|\omega ,o\rangle \,  \label{2.1}
\end{equation}
Then, $H$ and $O$ can be expressed as:

\begin{equation}
H{\text{ }}=\int_{0}^{\infty }\int_{o}\omega \;|\omega ,o\rangle \langle
\omega ,o|\;d\omega \,do\qquad ,\qquad O=\int_{0}^{\infty
}\int_{o}o\;|\omega ,o\rangle \langle \omega ,o|\;d\omega \,do  \label{2.2}
\end{equation}
In addition to $H$ and $O$, there are additional observables that may or may
not commute with $H$ and $O$. Then, a generic observable $A$ has the
following form:

\begin{equation}
A=\int_{0}^{\infty }\int_{0}^{\infty }\int_{o}\int_{o^{\prime }}A(\omega
,\omega ^{\prime },o,o^{\prime })\;|\omega ,o\rangle \langle \omega ^{\prime
},o^{\prime }|\;d\omega \,d\omega ^{\prime }\,do\,do^{\prime }\,  \label{2.3}
\end{equation}
where $A(\omega ,\omega ^{\prime },o,o^{\prime })$ could be, in principle, a
distributional kernel. However, we will not work with the set of all the
possible observables of the system, but only with a subset of it. The
condition that defines this subset is given by the choice of the kernel $%
A(\omega ,\omega ^{\prime },o,o^{\prime })$, which it is usually taken to be
(\cite{AS3}, \cite{ASLT}, \cite{Petrus}):

\begin{equation}
A(\omega ,\omega ^{\prime },o,o^{\prime })=A(\omega ,o,o^{\prime })\;\delta
(\omega -\omega ^{\prime })+A(\omega ,\omega ^{\prime },o,o^{\prime })\,
\label{2.4}
\end{equation}
where $A(\omega ,o,o^{\prime })$ and $A(\omega ,\omega ^{\prime
},o,o^{\prime })$ are sufficiently regular functions (see \cite{CG} for
details). Then, we will work with observables whose generic form is: 
\begin{equation}
A=\int_{0}^{\infty }\int_{o}\int_{o^{\prime }}A(\omega ,o,o^{\prime
})\;|\omega ,o,o^{\prime })\;d\omega \,do\,do^{\prime }\,+\int_{0}^{\infty
}\int_{0}^{\infty }\int_{o}\int_{o^{\prime }}A(\omega ,\omega ^{\prime
},o,o^{\prime })\;|\omega ,\omega ^{\prime },o,o^{\prime })\;d\omega
\,d\omega ^{\prime }\,do\,do^{\prime }  \label{2.5}
\end{equation}
where we have introduced $|\omega ,o,o^{\prime })=|\omega ,o\rangle \langle
\omega ,o^{\prime }|$ and $|\omega ,\omega ^{\prime },o,o^{\prime })=|\omega
,o\rangle \langle \omega ^{\prime },o^{\prime }|$. With the condition $%
\langle \omega ,o|\omega ^{\prime },o^{\prime }\rangle =\delta (\omega
-\omega ^{\prime })\,\delta (o-o^{\prime })$, the set of the operators of
the form (\ref{2.5}) is an algebra ${\cal A}$, and the observables are the
self-adjoint elements of ${\cal A}$ \cite{Petrus}, \cite{CG}\footnote{%
Although we will work with a subset of all the possible observables of the
system, the physical generality of the self-induced approach to decoherence
relies on the fact that the coordinates of the observables not belonging to $%
{\cal A}$ in the generalized eigenbasis of $\left\{ H,O\right\} $, being
singular, cannot be measured in laboratory and, therefore, they must always
be approximated by their averaged counterparts (for a full argument, see 
\cite{CL})}.

The first term of eq.(\ref{2.5}) represents the observables that commute
with those of the CSCO $\{H,O\}$, and it will be called the {\it singular
component} $A_{S}$ of $A$; the second term of eq.(\ref{2.5}) will be called
the {\it regular component} $A_{R}$ of $A$: 
\begin{equation}
A{_{S}}:=\int_{o}\int_{o^{\prime }}\int_{0}^{\infty }A(\omega ,o,o^{\prime
})\;|\omega ,o,o^{\prime })\;d\omega \,do\,do^{\prime }\quad ;\quad {A_{R}}%
:=\int_{o}\int_{o^{\prime }}\int_{0}^{\infty }\int_{0}^{\infty }A(\omega
,\omega ^{\prime },o,o^{\prime })\;|\omega ,\omega ^{\prime },o,o^{\prime
})\;d\omega \,d\omega ^{\prime }\,do\,do^{\prime }  \label{2.6}
\end{equation}
The operators $A_{S}$ and $A_{R}$ form the singular and regular algebras $%
{\cal A}_{S}$ and ${\cal A}_{R}$ respectively: $A{_{S}}\in {\cal A}_{S}$ and 
$A_{R}\in {\cal A}_{R}$. The generalized eigenbases of ${\cal A}_{S}$ and $%
{\cal A}_{R}$ are $\{|\omega ,o,o^{\prime })\}$ and $\{|\omega ,\omega
^{\prime },o,o^{\prime })\}$ since they span the algebras ${\cal A}_{S}$ and 
${\cal A}_{R}$ respectively. Note that these two algebras have a trivial
intersection and that the algebra ${\cal A}$ is the direct sum of both: $%
{\cal A}={\cal A}_{S}\oplus {\cal A}_{R}$ \cite{CG}.

\subsection{States as linear functionals}

States are continuous linear functionals on the algebra ${\cal A}$ defined
as above. Let ${\cal A}^{*}$ be the dual space of ${\cal A}$ (the vector
space of all linear continuous functionals on ${\cal A}$). In our notation,
the action of the functional $\rho \in {\cal A}^{*}$ onto the operator $A\in 
{\cal A}$ is denoted as $(\rho |A)$. With this notation we define $(\omega
,o,o^{\prime }|$ and $(\omega ,\omega ^{\prime },o,o^{\prime }|$ as:

\begin{equation}
(\omega ,o,o^{\prime }|A)=A(\omega ,o,o^{\prime })\qquad ,\qquad (\omega
,\omega ^{\prime },o,o^{\prime }|A)=A(\omega ,\omega ^{\prime },o,o^{\prime
})\,  \label{2.7}
\end{equation}
for all $A\in {\cal A}$. It can be shown that $(\omega ,o,o^{\prime }|$ and $%
(\omega ,\omega ^{\prime },o,o^{\prime }|$ are in ${\cal A}^{*}$ for all
values of $\omega $, $\omega ^{\prime }$, $o$, $o^{\prime }$ \cite{CG}. In
addition, if ${\cal A}_{S}^{*}$ is the dual of ${\cal A}_{S}$ and ${\cal A}%
_{R}^{*}$ is the dual of ${\cal A}_{R}$, it can be shown that $(\omega
,o,o^{\prime }|\in {\cal A}_{S}^{*}$ and $(\omega ,\omega ^{\prime
},o,o^{\prime }|\in {\cal A}_{R}^{*}$ \cite{Petrus}, \cite{CG}, and that the
following relations hold: 
\[
(\omega ,o,s|\omega ^{\prime },o^{\prime },s^{\prime })=\delta (\omega
-\omega ^{\prime })\delta (o-o^{\prime })\delta (s-s^{\prime
})\;\;\;,\;\;\;(\omega ,\sigma ,o,s|\omega ^{\prime },\sigma ^{\prime
},o^{\prime },s^{\prime })=\delta (\omega -\omega ^{\prime })\delta (\sigma
-\sigma ^{\prime })\delta (o-o^{\prime })\delta (s-s^{\prime })\,,
\]

\begin{equation}
(\omega ,\sigma ,|\omega ^{\prime },\sigma ^{\prime },o^{\prime },s^{\prime
})=(\omega ,\sigma ,o,s|\omega ^{\prime },\sigma ^{\prime })=0\,  \label{2.8}
\end{equation}
An element of the dual ${\cal A}^{*}$ can be expressed as:

\begin{equation}
{\rho }=\int_{0}^{\infty }\int_{o}\int_{o^{\prime }}\rho (\omega
,o,o^{\prime })\;(\omega ,o,o^{\prime }|\;d\omega \,do\,do^{\prime
}+\int_{0}^{\infty }\int_{0}^{\infty }\int_{o}\int_{o^{\prime }}\rho (\omega
,\omega ^{\prime },o,o^{\prime })\;(\omega ,\omega ^{\prime },o,o^{\prime
}|\;d\omega \,d\omega ^{\prime }\,do\,do^{\prime }  \label{2.9}
\end{equation}
Again, we will call the first term of eq.(\ref{2.9}) the {\it singular
component} $\rho _{S}$ of $\rho $, and the second term of eq.(\ref{2.9}) the 
{\it regular component} $\rho _{R}$ of $\rho $, where $\rho _{S}\in {\cal A}%
_{S}^{*}$ and $\rho _{R}\in {\cal A}_{R}^{*}$: 
\begin{equation}
{\rho }_{S}:=\int_{0}^{\infty }\int_{o}\int_{o^{\prime }}\rho (\omega
,o,o^{\prime })\,(\omega ,o,o^{\prime }|\,d\omega \,do\,do^{\prime }\quad
;\quad {\rho }_{R}:=\int_{0}^{\infty }\int_{0}^{\infty
}\int_{o}\int_{o^{\prime }}\rho (\omega ,\omega ^{\prime },o,o^{\prime
})\,(\omega ,\omega ^{\prime },o,o^{\prime }|\,d\omega \,d\omega ^{\prime
}\,do\,do^{\prime }  \label{2.10}
\end{equation}
The action of the functional $\rho $ on the operator $A$ is given by:

\begin{equation}
(\rho |A)=\int_{0}^{\infty }\int_{o}\int_{o^{\prime }}\rho (\omega
,o,o^{\prime })\,A(\omega ,o,o^{\prime })\,d\omega \,do\,do^{\prime
}+\int_{0}^{\infty }\int_{0}^{\infty }\int_{o}\int_{o^{\prime }}\rho (\omega
,\omega ^{\prime },o,o^{\prime })\,A(\omega ,\omega ^{\prime },o,o^{\prime
})\,d\omega \,d\omega ^{\prime }\,do\,do^{\prime }  \label{2.11}
\end{equation}
It is interesting to remark that, although $A(\omega ,o,o^{\prime })$ and $%
A(\omega ,\omega ^{\prime },o,o^{\prime })$ must be regular, well behaved
functions (polynomials and Schwartz functions, see \cite{CG}), this is not
the case of $\rho (\omega ,o,o^{\prime })$ and $\rho (\omega ,\omega
^{\prime },o,o^{\prime })$, which may be singular. For instance, the
functional $(\eta ,s,s^{\prime }|$ can be written in the form (\ref{2.9})
with $\rho (\omega ,o,o^{\prime })=\delta (\eta -\omega )\,\delta
(s-o)\,\delta (s^{\prime }-o^{\prime })$ and $\rho (\omega ,\omega ^{\prime
},o,o^{\prime })=0$.

The condition of positivity for a functional $f$ means that, if $f\in {\cal A%
}^{*}$ and $A\in {\cal A}$, then $f(A^{\dagger }A)\ge 0$, where $A^{\dagger }
$ is the adjoint of $A$. In our case we will require positivity to ${\rho }%
_{S}$, and this implies that: 
\begin{equation}
\rho (\omega ,o,o^{\prime })\geq 0\,.  \label{2.12}
\end{equation}
The condition of normalization for a functional $f$ means that the identity $%
I$ must be an element of the algebra and that $f(I)=1$. In our case, we will
normalize only $\rho _{S}$\footnote{%
We require positivity and normalization for $\rho _{S}$ since, as we will
see, it is the only component of $\rho $ that remains after decoherence. In
addition, $(\rho _{R}|I)=0$ for any  $\rho _{R}\in {\cal A}_{R}^{*}$.}: 
\begin{equation}
I=\int_{0}^{\infty }\int_{o}|\omega ,o\rangle \langle \omega ,o|\,\,d\omega
\,do\Longrightarrow (\rho _{S}|I)=\int_{0}^{\infty }\int_{o}\int_{o^{\prime
}}\rho (\omega ,o,o^{\prime })\,d\omega \,do\,do^{\prime }=1  \label{2.13}
\end{equation}
Finally, note that ${\cal A}^{*}={\cal A}_{S}^{*}\oplus {\cal A}_{R}^{*}$ 
\cite{Petrus}, and that $\{(\omega ,o,o^{\prime }|\}$ is a generalized basis
of ${\cal A}_{S}^{*}$ and $\{(\omega ,\omega ^{\prime },o,o^{\prime }|\}$ is
a generalized basis for ${\cal A}_{R}^{*}$.

\section{Self-induced decoherence}

\subsection{Decoherence in energy}

Let us now consider the time evolution of the system. Since $\rho $ is a
functional, its time evolution in the Schr\"{o}dinger picture cannot be
directly computed by means of the Liouville-von Neumann equation.
Nevertheless, this equation also describes the time evolution of the
observable $A$ in the Heisenberg picture: 
\begin{equation}
i\hbar \partial _{t}{A}=-[{H},{A}]={\Bbb L}A\Rightarrow {A}(t)=\exp (it\,%
{\Bbb L}{/\hbar )}\,{A}(0)  \label{3.1}
\end{equation}
where $H$ is the Hamiltonian that governs the time evolution, and ${\Bbb L}$
is the Liouville operator associated to the Hamiltonian $H$: ${\Bbb L}%
\,A=-[H,A]$. Once the time evolution of $A$ has been computed, the time
evolution of $\rho $ can be obtained by means of the {\it duality formula}:

\begin{equation}
(\rho |\exp (-it\,{\Bbb L}{/\hbar )}\,{A})=(\exp (it\,{\Bbb L}{/\hbar )}\rho
|A)  \label{3.2}
\end{equation}
This equation gives the time evolution of $\rho $, which satisfies the
Liouville-von Neumann equation: 
\begin{equation}
i\hbar \partial _{t}{\rho }=[{H},{\rho }]=-{\Bbb L}\rho \Rightarrow \rho
(t)=\exp (it\,{\Bbb L}{/\hbar )}\rho (0)  \label{3.3}
\end{equation}

In order to follow this strategy in our case, we begin by applying the
Liouville-von Neumann evolution equation to the generalized basis $\{|\omega
,o,o^{\prime })=|\omega ,o\rangle \langle \omega ,o^{\prime }|$, $|\omega
,\omega ^{\prime },o,o^{\prime })=|\omega ,o\rangle \langle \omega ^{\prime
},o^{\prime }|\}$. Since $H|\omega ,o,o^{\prime }\rangle =\omega \,|\omega
,o,o^{\prime }\rangle $, we have that: 
\begin{eqnarray}
{\Bbb L}|\omega ,o,o^{\prime }) &=&-H|\omega ,o\rangle \langle \omega
,o^{\prime }|+|\omega ,o\rangle \langle \omega ,o^{\prime }|H=-(\omega
-\omega )|\omega ,o,o^{\prime })=0  \label{3.4} \\
{\Bbb L}|\omega ,\omega ^{\prime },o,o^{\prime }) &=&-H|\omega ,o\rangle
\langle \omega ^{\prime },o^{\prime }|+|\omega ,o\rangle \langle \omega
^{\prime },o^{\prime }|H=-(\omega -\omega ^{\prime })|\omega ,\omega
^{\prime },o,o^{\prime })  \label{3.5}
\end{eqnarray}
This means that the generalized basis $\{|\omega ,o,o^{\prime })$, $|\omega
,\omega ^{\prime },o,o^{\prime })\}$ is an eigenbasis of the operator ${\Bbb %
L}$. Moreover, ${\Bbb L}|\omega ,o,o^{\prime })=0$ implies that not only the 
$|\omega ,o,o^{\prime })$, but also all the singular operators $A_{S}\in 
{\cal A}_{S}$ are time invariant ($e^{-it\,{\Bbb L}{/\hbar }}A_{S}=A_{S}$),
since:

\begin{equation}
{\Bbb L}A_{S}={\Bbb L}\int_{0}^{\infty }\int_{o}\int_{o^{\prime }}A(\omega
,o,o^{\prime })\,|\omega ,o,o^{\prime })\,d\omega \,do\,do^{\prime
}=\int_{0}^{\infty }\int_{o}\int_{o^{\prime }}A(\omega ,o,o^{\prime })\,(%
{\Bbb L}\,|\omega ,o,o^{\prime }))\,d\omega \,do\,do^{\prime }=0  \label{3.6}
\end{equation}
Therefore, for any $A\in {\cal A}$, ${\Bbb L}A={\Bbb L}A_{S}+{\Bbb L}A_{R}=%
{\Bbb L}A_{R}$. Moreover, due to eq.(\ref{3.5}) ${\Bbb L}A_{R}\in {\cal A}%
_{R}$.

From eq.(\ref{3.5}), it can be obtained:

\begin{equation}
e^{-it\,{\Bbb L}{/\hbar }}\,|\omega ,\omega ^{\prime },o,o^{\prime
})=e^{it(\omega -\omega ^{\prime })/\hbar }\,|\omega ,\omega ^{\prime
},o,o^{\prime })\,  \label{3.7}
\end{equation}
and, hence, for any $A_{R}\in {\cal A}_{R}$ we have:

\begin{equation}
e^{-it\,{\Bbb L}{/\hbar }}\,A_{R}=\int_{o}\int_{o^{\prime }}\int_{0}^{\infty
}\int_{0}^{\infty }A(\omega ,\omega ^{\prime },o,o^{\prime })\,e^{it(\omega
-\omega ^{\prime })/\hbar }\,|\omega ,\omega ^{\prime },o,o^{\prime
})\,d\omega \,d\omega ^{\prime }\,do\,do^{\prime }  \label{3.8}
\end{equation}
Then, for any $A\in {\cal A}$, we obtain the following time evolution in the
Heisenberg picture:

\begin{equation}
e^{-it\,{\Bbb L}{/\hbar }}\,A=\int_{0}^{\infty }\int_{o}\int_{o^{\prime
}}A(\omega ,o)|\omega ,o,o^{\prime })\,d\omega \,do\,do^{\prime
}+\int_{0}^{\infty }\int_{0}^{\infty }\int_{o}\int_{o^{\prime }}A(\omega
,\omega ^{\prime },o,o^{\prime })\,e^{it(\omega -\omega ^{\prime })/\hbar
}\,|\omega ,\omega ^{\prime },o,o^{\prime })\,d\omega \,d\omega ^{\prime
}\,do\,do^{\prime }\,  \label{3.9}
\end{equation}
A similar situation arises when we consider the time evolution of the
states. Since $|\omega ,o,o^{\prime })$ is time invariant, the duality
formula tells us that the functionals $(\omega ,o,o^{\prime }|$ are time
invariant; therefore, all the singular functionals $\rho _{S}\in {\cal A}%
_{S}^{*}$ are also time invariant.

Now we can compute $(\rho |A(t))$ which, by the duality formula (\ref{3.2}),
is equal to $(\rho (t)|A)$: 
\begin{eqnarray}
(\rho |A(t)) &=&(\rho (t)|A)=\int_{0}^{\infty }\int_{o}\int_{o^{\prime
}}\rho (\omega ,o,o^{\prime })\,\,A(\omega ,o,o^{\prime })\,d\omega
\,do\,do^{\prime }+  \nonumber \\
&&\int_{0}^{\infty }\int_{0}^{\infty }\int_{o}\int_{o^{\prime }}\rho (\omega
,\omega ^{\prime },o,o^{\prime })\,e^{i(\omega -\omega ^{\prime })t/\hbar
}A(\omega ^{\prime },\omega ,o^{\prime },o)\,d\omega \,d\omega ^{\prime
}\,do\,do^{\prime }  \label{3.10}
\end{eqnarray}
Considering that $A$ is arbitrary, we obtain the final equation for the
evolution of the functional $\rho $ as: 
\begin{equation}
{\rho }(t)=\int_{0}^{\infty }\int_{o}\int_{o^{\prime }}\rho (\omega
,o,o^{\prime })(\omega ,o,o^{\prime }|\,d\omega \,do\,do^{\prime
}+\int_{0}^{\infty }\int_{0}^{\infty }\int_{o}\int_{o^{\prime }}\rho (\omega
,\omega ^{\prime },o,o^{\prime })e^{i(\omega -\omega ^{\prime })t/\hbar
}(\omega ,\omega ^{\prime },o,o^{\prime }|\,d\omega \,d\omega ^{\prime
}\,do\,do^{\prime }\,  \label{3.11}
\end{equation}
where we will call the first term ``invariant part'' and the second term
``fluctuating part'' of ${\rho }(t)$.

If we now consider the states $\rho $ such that the product $\rho (\omega
,\omega ^{\prime },o,o^{\prime })\,A(\omega ,\omega ^{\prime },o,o^{\prime
}) $ is integrable, the Riemann-Lebesgue theorem \cite{RSII} can be applied
to eq.(\ref{3.10}) to conclude that:

\begin{equation}
\lim_{t\rightarrow \infty }({\rho }(t)|A)=\int_{0}^{\infty
}\int_{o}\int_{o^{\prime }}\rho (\omega ,o,o^{\prime })A(\omega ,o,o^{\prime
})\,d\omega \,do\,do^{\prime }=({\rho }_{*}|A)\qquad \text{{\it for any} }%
A\in {\cal A}\,  \label{3.12}
\end{equation}
where the functional $\rho _{*}$ is precisely the singular component $\rho
_{S}$ of $\rho $ (see eq.(\ref{2.10})):

\begin{equation}
{\rho }_{*}=\int_{0}^{\infty }\int_{o}\int_{o^{\prime }}\rho (\omega
,o,o^{\prime })(\omega ,o,o^{\prime }|\,d\omega do\,do^{\prime }\,
\label{3.13}
\end{equation}

The physical meaning of this process can be understood when we consider that
the mean value of the observable $A$ in the state $\rho $ can be computed as 
$\langle A\rangle _{\rho }=({\rho }|A)$. Therefore, eq.(\ref{3.12}) can be
rewritten as: 
\begin{equation}
\lim_{t\rightarrow \infty }\langle A\rangle _{\rho (t)}=\langle A\rangle
_{\rho _{*}}\qquad \text{{\it for any} }A\in {\cal A}\,  \label{3.14}
\end{equation}
Of course, this limit does not contradict the fact that the off-diagonal
terms of a functional $\rho $ representing the quantum state of a closed
system never vanish through the {\it unitary} evolution described by the
Liouville-von Neumann equation. What self-induced decoherence shows is that
the mean value $\langle A\rangle _{\rho (t)}$ of any observable $A\in {\cal A%
}\,$ will evolve in such a way that, for $t\rightarrow \infty $, it can be
computed {\it as if }the system were in a state ${\rho }_{*}$ where the
off-diagonal terms have vanished. Formally this is expressed by the fact
that, although we strictly obtain the limit (\ref{3.14}) (or (\ref{3.12})),
the state $\rho (t)$ has only a {\it weak limit}:

\begin{equation}
w-\lim_{t\rightarrow \infty }{\rho }(t)={\rho }_{*}  \label{3.15}
\end{equation}
This weak limit means that, even if $\rho (t)$ always follows a unitary
evolution, the system decoheres {\it from an observational point of view},
that is, from the viewpoint given by the observable $A$, for any $A\in {\cal %
A}\,$.

\subsection{Decoherence in the remaining variables}

As we have seen, for $t\rightarrow \infty $ the system decoheres in energy
since ${\rho }_{*}$ turns out to be diagonal in $\omega $. However, we would
like to obtain a state diagonal in all the variables. To the extent that we
have taken the limit $t\rightarrow \infty $, it is impossible that a new
process diagonalizes the $o-$variables. As we will see, when a convenient
basis is chosen, the diagonalization of $\rho $ can be completed. This
second stage necessarily depends on the initial condition ${\rho }$ at $t={0}
$, since ${\rho }_{*}$ is a constant of motion.

Let us consider a unitary operator $U$ that keeps the Hamiltonian invariant
but changes the set of observables $\{O_{1},O_{2},\dots ,O_{N}\}$ into the
set $\{P_{1},P_{2},\dots ,P_{N}\}$, where $\{H,P_{1},P_{2},\dots ,P_{N}\}$
is also a CSCO. The simplest form of $U$ is given by:

\begin{equation}
U=\int_{0}^{\infty }\int_{p}\int_{o}U(\omega ,p,o)\,|\omega ,p\rangle
\langle \omega ,o|\,d\omega \,dp\,do\,  \label{3.16}
\end{equation}
The action of $U$ on the ket $|\omega ,o\rangle $ defines the action of $U$
on any ket, since the kets $|\omega ,o\rangle $ belong to a generalized
basis. This action can be easily computed as:

\begin{equation}
|\omega ,p\rangle :=U\,|\omega ,o\rangle =\int_{p}U(\omega ,p,o)\,|\omega
,o\rangle \,dp\,  \label{3.17}
\end{equation}
The unitarity of $U$ implies that $UU^{-1}=I$. From here, we obtain:

\begin{equation}
\int_{o}U(\omega ,p,o)\,U^{*}(\omega ,p^{\prime },o)\,do=\delta (p-p^{\prime
})\,  \label{3.18}
\end{equation}
In the new representation, the operator $A$ takes the form:

\begin{equation}
A=\int_{0}^{\infty }\int_{p}\int_{p^{\prime }}A(\omega ,p,p^{\prime
})\,|\omega ,p\rangle \langle \omega ,p^{\prime }|\,d\omega \,dp\,dp^{\prime
}.  \label{3.19}
\end{equation}
Introducing eq.(\ref{3.17}) into eq.(\ref{3.19}) gives:

\begin{equation}
A=\int_{0}^{\infty }\int_{p}\int_{p^{\prime }}\int_{o}\int_{o^{\prime
}}U(\omega ,p,o)\,\,A(\omega ,p,p^{\prime })\,U^{*}(\omega ,p^{\prime
},o^{\prime })\,|\omega ,o\rangle \langle \omega ,o^{\prime }|\,d\omega
\,dp\,dp^{\prime }\,do\,do^{\prime }\,  \label{3.20}
\end{equation}
Therefore, the coordinates of $A$ in the old basis are:

\begin{equation}
A(\omega ,o,o^{\prime })=\int_{p}\int_{p^{\prime }}U(\omega
,p,o)\,\,A(\omega ,p,p^{\prime })\,U^{*}(\omega ,p^{\prime },o^{\prime
})\,dp\,dp^{\prime }\,  \label{3.21}
\end{equation}
Since $U$ is a unitary operator, eq.(\ref{3.21}) is invertible:

\begin{equation}
A(\omega ,p,p^{\prime })=\int_{o}\int_{o^{\prime }}U(\omega ,p^{\prime
},o^{\prime })\,\,A(\omega ,o,o^{\prime })\,U^{*}(\omega
,p,o)\,do\,do^{\prime }\,  \label{3.22}
\end{equation}
and, finally, by duality one finds that:

\begin{equation}
\rho (\omega ,p,p^{\prime })=\int_{o}\int_{o^{\prime }}U(\omega ,p^{\prime
},o^{\prime })\,\rho (\omega ,o,o^{\prime })\,U^{*}(\omega
,p,o)\,do\,do^{\prime }  \label{3.23}
\end{equation}

If $\rho =\int_{0}^{\infty }\int_{o}\int_{o^{\prime }}\rho (\omega
,o,o^{\prime })\,|\omega ,o\rangle \langle \omega ,o^{\prime }|\,d\omega
\,do\,do^{\prime }$ is a state, it must be positively defined and self
adjoint. This implies that $\rho (\omega ,o,o^{\prime })=\rho ^{*}(\omega
,o^{\prime },o)$ \cite{CG}. Then, we can choose $U(\omega ,p,o)$ such that:

\begin{equation}
\rho (\omega ,p,p^{\prime })=\rho (\omega ,p)\,\delta (p-p^{\prime })
\label{3.24}
\end{equation}
and this completes the diagonalization. In the new basis, eqs.(\ref{3.12})
and (\ref{3.13}) of the previous subsection become:

\begin{equation}
\lim_{t\rightarrow \infty }({\rho }(t)|A)=\int_{p}\int_{0}^{\infty }\rho
(\omega ,p)A(\omega ,p)\,d\omega \,dp=({\rho }_{*}|A)\qquad \text{{\it for
any} }A\in {\cal A}\,\,  \label{3.25}
\end{equation}

\begin{equation}
{\rho }_{*}=\int_{p}\int_{0}^{\infty }\rho (\omega ,p)(\omega ,p|\,d\omega
\,dp  \label{3.26}
\end{equation}
where now ${\rho }_{*}$ is completely diagonal in $\omega $ and $p$. The
generalized basis given by $\{|\omega ,p,p^{\prime }),\,|\omega ,\omega
^{\prime },p,p^{\prime })\}$ is the preferred basis (also called ''pointer
basis''\footnote{%
We prefer to use the term ''preferred basis'' instead of ''pointer basis''
since it arises not only in measurements contexts where the pointer of a
measuring device is involved.}), as presented in \cite{Deco} and extensively
discussed in \cite{CL}. On the other hand, the decoherence time $t_{D}$ can
be computed\footnote{%
Of course, in the case where $t_{D}\rightarrow \infty $, the system does not
decohere.} as in paper \cite{Rolo}.

\section{The Wigner transformation}

\subsection{Characterization of the Wigner transformation}

The Weyl transformation maps functions or generalized functions on phase
space into operators \cite{DS}. Thus, the Wigner transformation maps
operators into functions on the phase space \cite{Wigner}, \cite{Symb}. If $%
A $ is an operator, we will denote the function corresponding to $A$ via the
Wigner transformation by $symbA$ or $A(\phi )$, where $\phi =({\bf q},{\bf p}%
)=(q_{1},q_{2},\dots ,q_{N+1},p_{1},p_{2},\dots ,p_{N+1})$ is a point in
flat phase space. The function $A(\phi )=symbA$ is called {\it Wigner symbol}
or {\it Wigner function} of the operator $A$. As the Weyl transformation is
a one to one mapping, the image of the algebras ${\cal A}$, ${\cal A}_{S}$
and ${\cal A}_{R}$ by the Wigner transformation are non-commutative algebras
of functions denoted by ${\cal L}$, ${\cal L}_{S}$ and ${\cal L}_{R}$
respectively, where ${\cal L}={\cal L}_{S}\oplus {\cal L}_{R}$. Since the
image of $A$ by the Wigner transformation is the function $symbA$, it seems
quite natural to use the notation $symb$ to denote the Wigner transformation
itself, so that:

\begin{equation}
symb:{\cal A}\longmapsto {\cal L}\qquad ,\qquad symb:{\cal A}_{R}\longmapsto 
{\cal L}_{R}\qquad ,\qquad symb:{\cal A}_{S}\longmapsto {\cal L}_{S}
\label{4.1}
\end{equation}

Let us define the mapping ${\cal A}_{R}\longmapsto {\cal L}_{R}$ for regular
observables as usual \cite{Symb}. First, consider the phase space (in this
case\footnote{%
The fact that the dimension of the phase space is $2(N+1)$, where $N+1$ is
the number of observables of the CSCO, amounts to the integrability of the
classical system resulting from the Wigner transformation. Non-integrable
cases will be considered elsewhere.} ${\Bbb R}^{2({N+1)}}$) and endow it
with the symplectic form:

\begin{equation}
\omega _{ab}=\left( 
\begin{array}{ll}
\text{ }0 & I_{N+1} \\ 
-I_{N+1} & 0
\end{array}
\right) \qquad \omega ^{ab}=\left( 
\begin{array}{ll}
0 & -I_{N+1} \\ 
I_{N+1} & \text{ }0
\end{array}
\right)  \label{4.2}
\end{equation}
Then, let $\widehat{f}$ be an operator such that $symb\widehat{f}=f(\phi )$.
This transformation is defined by the usual Wigner recipe as:

\begin{equation}
symb\widehat{f}=f(\phi ):=\int d^{2{N+1}}\psi \exp \left( \frac{i}{\hbar }%
\psi ^{a}\omega _{ab}\phi ^{b}\right) Tr\left( {T}(\psi )\widehat{f}\right)
\,  \label{4.3}
\end{equation}
where $\psi ^{a}$ and $\phi ^{b}$ denote the $a$-th and the $b$-th
components of the points $\psi $ and $\phi $ on phase space, respectively.
Here:

\begin{equation}
{T}(\psi )=\exp \left( \frac{i}{\hbar }\psi ^{a}\omega _{ab}\widehat{{\phi }%
^{b}}\right)  \label{4.4}
\end{equation}
where:

\begin{equation}
\widehat{\phi }=(\widehat{q}_{1},\dots ,\widehat{q}_{N+1},-i\hbar \frac{%
\partial }{\partial q_{1}},\dots ,-i\hbar \frac{\partial }{\partial q_{N+1}})
\label{4.5}
\end{equation}
and $\widehat{q}_{i}$, $i=1,2,\dots ,N+1$, is the $i$-th component of the
position operator on the Hilbert space $L^{2}({\Bbb R}^{N+1})$.

The non-commutative product that corresponds to the product of operators in $%
{\cal L}$ (or ${\cal L}_{S}$ and ${\cal L}_{R}$) is called {\it star product}%
, and it is given by:

\begin{equation}
symb(\widehat{f}\widehat{g})=symb\widehat{f}*symb\widehat{g}=(f*g)(\phi )\,
\label{4.6}
\end{equation}
where $f(\phi )$ and $g(\phi )$ are the Wigner symbols of the operators $%
\widehat{f}$ and $\widehat{g}$, respectively. It can be proven (\cite{Wigner}%
, eq.(2.59); for more general expansions, see \cite{Ruso}) that:

\begin{equation}
(f*g)(\phi )=f(\phi )\exp \left( \frac{i\hbar }{2}\overleftarrow{\partial }%
_{a}\omega ^{ab}\overrightarrow{\partial }_{b}\right) g(\phi )=g(\phi )\exp
\left( -\frac{i\hbar }{2}\overleftarrow{\partial }_{a}\omega ^{ab}%
\overrightarrow{\partial }_{b}\right) f(\phi )\,  \label{4.7}
\end{equation}
The {\it Moyal bracket} is the Wigner symbol corresponding to the commutator
in ${\cal L}$:

\begin{equation}
\{f,g\}_{mb}=\frac{1}{i\hbar }(f*g-g*f)=symb\left( \frac{1}{i\hbar }%
[f,g]\right)  \label{4.8}
\end{equation}
Then, if we expand the last two equations in power series of $\hbar $, we
obtain \cite{DS}:

\begin{equation}
(f*g)(\phi )=f(\phi )g(\phi )+\sum_{r=1}\hbar ^{r}P^{r}(f(\phi )g(\phi ))
\label{4.9}
\end{equation}

\begin{equation}
\{f,g\}_{mb}=\{f,g\}_{pb}+\sum_{r=1}\hbar ^{2r}P^{2r+1}(f(\phi )g(\phi ))
\label{4.10}
\end{equation}
where the $P^{r}$ are the coefficients obtained by means of eq.(\ref{4.7})
and $pb$ means Poisson bracket. This suggests that, in the limit $\hbar
\rightarrow 0$, the star product should become the ordinary product and the
Moyal bracket should become the Poisson bracket. In fact, this is the case
in many circumstances although not in all, because in some cases the
coefficients $P^{r}$ may contain factors of the form $\hbar ^{-1}$, making
the limit $\hbar \rightarrow 0$ singular (see \cite{DS}, \cite{Ballentine});
in those cases, the problem 3 mentioned in the Introduction arises. From eq.(%
\ref{4.7}) we see that factors $\hbar ^{-1}$ can only come from the symbols $%
f(\phi )$ or $g(\phi )$; then, if these functions do not depend on $\hbar
^{-1}$, the limit $\hbar \rightarrow 0$ is regular and can be considered as
the proper macroscopic limit.

Finally, let us observe that if $\widehat{f}$ commute with $\widehat{g}$,
eqs.(\ref{4.7}) and (\ref{4.9}) become:

\begin{equation}
(f*g)(\phi )=f(\phi )\cos \left( -\frac{i\hbar }{2}\overleftarrow{\partial }%
_{a}\omega ^{ab}\overrightarrow{\partial }_{b}\right) g(\phi )  \label{4.11}
\end{equation}
and, hence, in the simple cases with no factors $\hbar ^{-1}$ we obtain:

\begin{equation}
(f*g)(\phi )=f(\phi )g(\phi )+0(\hbar ^{2})  \label{4.12}
\end{equation}

\subsection{The Wigner transformation of observables and states}

\smallskip In the previous subsection we have considered the Wigner
transformation for regular observables, as it is usually defined. But the
Wigner transformation has not been defined when singular distributions are
involved; therefore, the transformation must be defined in this case.

Let us go back to the algebra ${\cal A}_{S}$ of the observables that commute
with the CSCO $\{H,P_{1},\dots ,P_{N}\},$ also denoted by $\{H,P\}$ for
simplicity. An element of this algebra is given by:

\begin{equation}
A_{S}=\int_{0}^{\infty }\int_{p}A(\omega ,p)\,|\omega ,p)\,d\omega \,dp\,
\label{4.13}
\end{equation}
where $A(\omega ,p)$ is a regular function on its variables. The functional
calculus gives:

\begin{equation}
A_{S}=A(H,P)\,  \label{4.14}
\end{equation}
where $P:=(P_{1},P_{2},\dots ,P_{N})$. Since we are assuming that the
observables $H$ and $P_{i}$ are Weyl observables, they have Wigner functions 
$H(\phi )$ and $P_{i}(\phi )$, $i=1,2,\dots ,N$, respectively, not depending
on $\hbar ^{-1}$. As a consequence, if $\widehat{f}$ and $\widehat{g}$ are
any powers of $H$ or $P_{i}$, eq.(\ref{4.12}) holds. Therefore, if we also
assume that $A(\omega ,p)$ is analytic on its variables, we have that:

\begin{equation}
symbA_{S}=A_{S}(\phi )=A(H(\phi ),P(\phi ))+0(\hbar ^{2})  \label{4.15}
\end{equation}
This means that the problem 3 mentioned in the Introduction does not arise
in the singular algebra ${\cal A}_{S}$ when we work with Weyl operators. As
a consequence, the Wigner symbol of any observable $A\in $ ${\cal A}_{S}$,
when $\hbar \rightarrow 0$, is $A(H(\phi ),P(\phi ))$.

Now let us study the Wigner transformation for states of ${\cal A}^{*}$.
There are two cases that we have to consider. The first case is given by the
states in ${\cal A}^{*}$ that can be written as regular density operators.
These states are characterized by $\rho (\omega ,\omega ,p,p^{\prime })=\rho
(\omega ,p,p^{\prime })$ (see \cite{AS3}, \cite{ASLT}, \cite{Pra}, \cite
{Petrus}, \cite{CG}). Regular density operators have well defined Wigner
functions \cite{Symb}: we only must add a $(2\pi \hbar )^{N+1}$ factor to
eq.(\ref{4.3}) in order to obtain the usual normalization for the Wigner
function. The second case includes any other possibility, i.e., density
operators for which either $\rho (\omega ,\omega ,p,p^{\prime })\ne \rho
(\omega ,p,p^{\prime })$ or $\rho (\omega ,\omega ,p,p^{\prime })$ or $\rho
(\omega ,p,p^{\prime })$ are defined only in a distributional sense and not
as regular functions. The question arises about whether the operators of
this second type do or do not have a well defined Wigner function. The
answer is given by the duality formula (\ref{3.2}). If $\rho \in {\cal A}^{*}
$ and $A\in {\cal A}$, we can define $symb\rho \equiv \rho (\phi )$ in such
a way that it satisfies:

\begin{eqnarray}
(symb{}\rho |symb{}A) &:&=(\rho |A)=\int_{0}^{\infty
}\int_{p}\int_{p^{\prime }}A(\omega ,p,p^{\prime })\,\rho (\omega
,p,p^{\prime })\,d\omega \,dp\,dp^{\prime }+  \nonumber \\[0.03in]
&&\int_{0}^{\infty }\int_{0}^{\infty }\int_{p}\int_{p^{\prime }}A(\omega
,\omega ^{\prime },p,p^{\prime })\,\rho (\omega ,\omega ^{\prime
},p,p^{\prime })\,d\omega \,d\omega ^{\prime }\,dp\,dp^{\prime }
\label{4.16}
\end{eqnarray}
As the integrals in (\ref{4.16}) are well defined, for any $\rho \in {\cal A}%
^{*}$, $symb{}\rho $ is also well defined and belongs to the dual space $%
{\cal L}^{*}$ of the algebra ${\cal L}$. Let us recall the decomposition $%
{\cal L}^{*}={\cal L}_{S}^{*}\oplus {\cal L}_{R}^{*}$ and the fact that the
operation $symb$ is a bijection: $symb\,{\cal A}_{S}^{*}\longmapsto {\cal L}%
_{S}^{*}$. Definition (\ref{4.16}) will allow us to obtain $symb{}\rho _{S}$
in the next section.

\section{The classical limit}

We have seen that, as the result of decoherence, the regular part $\rho _{R}$
of $\rho $ vanishes and only the singular part $\rho _{S}={\rho }_{*}$
remains (see eq.(\ref{3.26})): 
\begin{equation}
{\rho }_{*}=\rho _{S}=\int_{0}^{\infty }\int_{p}\rho (\omega ,p)(\omega
,p|\,d\omega \,dp  \label{5.1}
\end{equation}
Then, the problem is to find the classical distribution $\rho _{c}(\phi )$
resulting from applying the macroscopic limit $\hbar \rightarrow 0$ to the
Wigner transformation of ${\rho }_{*}=\rho _{S}$: 
\begin{equation}
\rho _{c}(\phi )=\lim_{\hbar \rightarrow 0}symb{}\rho _{S}=\lim_{\hbar
\rightarrow 0}\int_{0}^{\infty }\int_{p}\rho (\omega ,p)\,symb(\omega
,p|\,d\omega \,dp=\int_{0}^{\infty }\int_{p}\rho (\omega ,p)\,\lim_{\hbar
\rightarrow 0}\left[ symb(\omega ,p|\right] \,d\omega \,dp  \label{5.2}
\end{equation}

The first step consists in obtaining the limit of $symb|\omega ,p)$ for $%
\hbar \rightarrow 0$. From eqs.(\ref{4.13}) and (\ref{4.15}) we know that: 
\begin{equation}
\lim_{\hbar \rightarrow 0}[symbA_{S}]=A(H(\phi ),P(\phi ))=\lim_{\hbar
\rightarrow 0}\int_{0}^{\infty }\int_{p}A(\omega ,p)\,symb|\omega
,p)\,d\omega \,dp=\int_{0}^{\infty }\int_{p}A(\omega ,p)\,[\lim_{\hbar
\rightarrow 0}symb|\omega ,p)]\,d\omega \,dp  \label{5.3}
\end{equation}
But $A(H(\phi ),P(\phi ))$ can also be written as:

\begin{equation}
A(H(\phi ),P(\phi ))=\int_{0}^{\infty }\int_{p}A(\omega ,p)\ \delta (H(\phi
)-\omega )\,\delta (P(\phi )-p)\,dp\,d\omega \,  \label{5.4}
\end{equation}
By comparing eq.(\ref{5.3}) and eq.(\ref{5.4}) we can conclude that\footnote{%
In the case of discrete spectrum we would have $A(\omega ,p)=\delta (\omega
-\omega ^{\prime })\delta _{pp^{\prime }}^{N}$ instead of $A(\omega
,p)=\delta (\omega -\omega ^{\prime })\,\delta (p-p^{\prime })$. Then, we
would obtain $symb|\omega ^{\prime },p^{\prime })=\delta (H(\phi )-\omega
^{\prime })\delta _{P(\phi )p^{\prime }}^{N}$, where $\delta _{P(\phi
)p^{\prime }}^{N}$ is a $N$ Kronecker $\delta $.}: 
\begin{equation}
\lim_{\hbar \rightarrow 0}symb\,|\omega ,p)=\delta (H(\phi )-\omega
)\,\delta (P(\phi )-p)  \label{5.5}
\end{equation}

The second step begins by remembering that, from eq.(\ref{4.16}) and eq.(\ref
{2.8}), $symb(\omega ,p|$ must satisfy: 
\begin{equation}
(symb(\,\omega ,p|\,|symb\,|\omega ^{\prime },p^{\prime }))=(\omega
,p|\omega ^{\prime },p^{\prime })=\delta (\omega -\omega ^{\prime })\,\delta
(p-p^{\prime })\,  \label{5.6}
\end{equation}
Since the r.h.s. of the last equation does not depend on $\hbar $, the limit
for $\hbar \rightarrow 0$ results: 
\begin{equation}
\lim_{\hbar \rightarrow 0}(symb(\,\omega ,p|\,|symb\,|\omega ^{\prime
},p^{\prime }))=(\lim_{\hbar \rightarrow 0}[symb(\,\omega ,p|]\,|\lim_{\hbar
\rightarrow 0}[symb\,|\omega ^{\prime },p^{\prime })])=\delta (\omega
-\omega ^{\prime })\,\delta (p-p^{\prime })  \label{5.7}
\end{equation}
We will use $\rho _{S\omega p}(\phi )$ to denote the limit for $\hbar
\rightarrow 0$ of $symb{}(\omega ,p|$: 
\begin{equation}
\rho _{S\omega p}(\phi ):=\lim_{\hbar \rightarrow 0}symb{}(\omega ,p|
\label{5.8}
\end{equation}
Therefore, replacing eq.(\ref{5.5}) and eq.(\ref{5.8}) into eq.(\ref{5.7})
we obtain: 
\begin{equation}
(\rho _{S\omega p}(\phi )\,\,|\,\delta (H(\phi )-\omega ^{\prime })\,\delta
(P(\phi )-p^{\prime }))=\delta (\omega -\omega ^{\prime })\,\delta
(p-p^{\prime })  \label{5.9}
\end{equation}

The final step consists in obtaining $\rho _{S\omega p}(\phi )$. Since we
have assumed that the number of operators in the CSCO $\{H,P_{1},\dots
,P_{N}\}$ coincides with the number of degrees of freedom of the system
under consideration and, as a consecuence, the phase space has dimension $%
2(N+1)$ (see footnote 9), then there exists a canonical transformation that
carries the position-momentum variables $\phi =({\bf q},{\bf p})$ into the
variables $\psi =(\tau (\phi ),\alpha _{1}(\phi ),\dots ,\alpha _{N}(\phi
),H(\phi ),P_{1}(\phi ),\dots ,P_{N}(\phi ))$, where $H(\phi )=symbH$, $%
P_{i}(\phi )=symb{}P_{i}$, $i=1,2,\dots ,N$, $\tau (\phi )$ is the conjugate
variable of $H(\phi )$, and the $\alpha _{i}(\phi )$ are the conjugate
variables of the $P_{i}(\phi )$. On the other hand, from subsection III.A we
know that all the singular operators $A{_{S}}\in {\cal A}_{S}$ and all the
singular functionals $\rho _{S}\in {\cal A}_{S}^{*}$ are time invariant.
Since the function $symb$ does not introduce time variables, all the $%
A_{S}(\phi )=symbA_{S}\in {\cal L}_{S}^{{}}$ and all the $\rho _{S}(\phi
)=symb\rho _{S}\in {\cal L}_{S}^{*}$ are also time invariant, and the same
holds for their limits for $\hbar \rightarrow 0$. Thus, if $A(\phi )\in 
{\cal L}_{S}$, $\rho (\phi )\in {\cal L}_{S}^{*}$, $A(\phi )=f(H(\phi
),P(\phi ))$, and $\rho (\phi )=g(H(\phi ),P(\phi ))$, then $(\rho (\phi
)|A(\phi ))$ can be expressed as:

\begin{equation}
(\rho (\phi )|A(\phi ))=\int_{\phi ^{2(N+1)}}\,f(H(\phi ),P(\phi
))\,\,g(H(\phi ),P(\phi ))\,\,d\phi ^{2(N+1)}=\int_{\tau }d\tau \int_{\alpha
}d\alpha \int_{H}\int_{P}\,f(H,P)\,\,g(H,P)\,dH\,dP  \label{5.10}
\end{equation}
We can call:

\begin{equation}
C(H,P)=\int_{{\cal M}(H,P)}d\tau \,d\alpha   \label{5.11}
\end{equation}
where $C(H,P)$ is the volume of the region ${\cal M}(H,P)$ of the
configuration manifold defined by the conditions $H=$ $const.$ and $P=$ $%
const.$ Then, eq.(\ref{5.10}) becomes:

\begin{equation}
(\rho (\phi )|A(\phi ))=C(H,P)\int_{H}\int_{P}\,f(H,P)\,\,g(H,P)\,dH\,dP
\label{5.12}
\end{equation}
In the simplest case of bounded integrable systems described by action-angle
variables, $C(H,P)$ is a constant equal to $(2\pi )^{N+1}$. Anyway, $C(H,P)$
is always a constant that we will ignore from now on in order to simplify
notation. If we now apply the result expressed by eq.(\ref{5.12}) to eq.(\ref
{5.9}), we obtain:

\begin{equation}
\int_{H}\int_{P}\rho _{S\omega p}(H,P)\,\delta (H-\omega ^{\prime })\,\delta
(P-p^{\prime })\,dH\,dP=\delta (\omega -\omega ^{\prime })\,\delta
(p-p^{\prime })\,  \label{5.13}
\end{equation}
This means that\footnote{%
In the discrete case, eq.(\ref{5.13}) reads:
\par
\[
\int_{H}\int_{P}\rho _{S\omega p}(\phi )\,\delta (H-\omega ^{\prime
})\,\delta _{Pp^{\prime }}^{N}\,dH\,dP^{N}=\delta (\omega -\omega ^{\prime
})\,\delta _{pp^{\prime }}^{N} 
\]
Then:
\par
\[
\rho _{S\omega ^{\prime }p^{\prime }}(\phi )=\delta (H(\phi )-\omega
^{\prime })\,\delta ^{N}(P(\phi )-p^{\prime })\, 
\]
}: 
\begin{equation}
\rho _{S\omega p}(H,P)=\delta (H-\omega )\,\delta (P-p)  \label{5.14}
\end{equation}
If we now go back to the variables $\phi $:

\begin{equation}
\rho _{S\omega p}(\phi )=\lim_{\hbar \rightarrow 0}symb{}(\omega ,p|=\delta
(H(\phi )-\omega )\,\delta (P(\phi )-p)  \label{5.15}
\end{equation}

Finally, we can obtain the classical distribution $\rho _{c}$ by replacing
the just obtained result (\ref{5.15}) into eq.(\ref{5.2}): 
\begin{equation}
\rho _{c}(\phi )=\int_{0}^{\infty }\int_{p}\rho (\omega ,p)\,\rho _{S\omega
p}(\phi )\,d\omega \,dp=\int_{0}^{\infty }\int_{p}\rho (\omega ,p)\,\delta
(H(\phi )-\omega )\,\delta (P(\phi )-p)\,d\omega \,dp  \label{5.16}
\end{equation}
As we can see, $\rho _{c}(\phi )$ is a {\it constant of motion}, as it was
expected due to the results obtained in Subsection III.A. Eq.(\ref{5.16})
has a clear{\it \ physical meaning}: $\rho _{c}(\phi )$ is a sum of
densities infinitely strongly peaked on the classical trajectories defined
by the constants of motion $H(\phi )=\omega $ and $P(\phi )=p$ and averaged
by the density function $\rho (\omega ,p)$ which is properly normalized
according to eq.(\ref{2.13}). As a consequence, $\rho _{c}(\phi )$ can be
conceived as sum of classical trajectories weighted by their corresponding
probabilities. This leads to the expected result: classical motion takes
place along a classical trajectory, and the probability of each possible
trajectory is given by the initial condition $\rho $ at $t=0$.

It is also interesting to consider the case where the initial condition $%
\rho $ at $t=0$ is such that the factor $\rho (\omega ,\omega ^{\prime
},p,p^{\prime })\,A(\omega ,\omega ^{\prime },p,p^{\prime })$ of eq.(\ref
{3.10}) is strongly peaked around $\omega -\omega ^{\prime }=E$. In this
case, the evolution factor $e^{i(\omega -\omega ^{\prime })t/\hbar }$ can be
approximated by $e^{iEt/\hbar }$. This shows two facts. First, there is an
interplay between the characteristic energy $E$ and the decoherence time:
the decoherence time becomes shorter as the energy is higher. Second, the
limit $E\rightarrow \infty $ plays the same mathematical role as $%
t\rightarrow \infty $, and represents the well known ''high energy limit'' 
\cite{Ballentine}: for high energies many systems behave in an almost
classical way, e.g., high energy orbits of atoms can be approximated by
classical trajectories\footnote{%
It is worth noting, however, that while $t$ is a perfectly well defined
variable, $E$ is just a characteristic energy that may be not well defined
in some cases. For this reason, the mathematically precise strategy is to
find the limit for $t\rightarrow \infty $ as we have done in the present
work.}.

In this section we have found the classical limit by applying the
macroscopic limit to the Wigner transformation of the state ${\rho }_{*}$
resulting from decoherence. Nevertheless, it is also possible to translate
the quantum evolution equation into the classical language via the Wigner
transformation in order to obtain the complete process of decoherence in
classical terms. This strategy leads to the same result as the one obtained
in the present section (see Appendix A).

Finally, let us recall the three problems of the traditional way of
conceiving the classical limit as they were presented in the Introduction,
and let us consider how and under what conditions they can be overcome from
the present approach:

\begin{enumerate}
\item  Although in general the Wigner function $\rho (\phi )$ of a state $%
\rho $ is not non-negatively defined, we can guarantee the non-negativeness
of $\rho _{c}(\phi )$. In fact, $\rho (\omega ,p)$ is non-negatively defined
due to its origin, since it represents the diagonal components of the
original state $\rho $ (it can be also proved that, in general, if $\rho
=\rho (\omega )=\rho (\omega ,\omega )$, then the limit for $\hbar
\rightarrow 0$ of $\rho (\phi )$ is non-negatively defined a.e.; see
Appendix B).

\item  Although only Hamiltonians of degree $\leq 2$ in $p$ and $q$ yield to
Hamiltonian fluxes that maintain the deformation invariant, this is not a
problem from the present perspective since, after decoherence, only the
singular algebra remains, and in this algebra both states and observables
are constants of motion.

\item  Although in some cases factors of the form $\hbar ^{-1}$ may appear
in the Wigner state function making the limit $\hbar \rightarrow 0$
singular, we have shown that this possibility is blocked when the
observables of the CSCO are Weyl observables. The requirement of working
with a CSCO consisting of Weyl observables is very weak since it does not
impose artificial constraints on the state $\rho $. Furthermore, to apply
the Weyl transformation to classical observables is the usual strategy in
practice for obtaining the corresponding quantum observables.
\end{enumerate}

\section{The physical meaning of the classical limit}

As we have seen, the classical limit of quantum mechanics involves two
elements:

\begin{enumerate}
\item  {\it Decoherence}: According to the self-induced approach,
decoherence is a physical process that depends on the own dynamics of a
closed quantum system governed by a Hamiltonian with continuous spectrum. As
the result of decoherence, in the infinite time limit the mean value of any
relevant observable can be computed as if the system were in the diagonal
state ${\rho }_{*}$. In other words, decoherence transforms standard
(non-Boolean) quantum mechanics into a Boolean quantum mechanics restricted
to states that are diagonal in the basis defined by the CSCO $\left\{
H,P\right\} $.

\item  {\it Macroscopicity}: For $\hbar \rightarrow 0$, the Wigner
transformation of the diagonal state ${\rho }_{*}$ turns out to be $\rho
_{c}(\phi )$, and it is resolved into an ensemble of classical trajectories
on phase space weighted by their corresponding probabilities. This means
that, in the macroscopic limit, the Wigner transformation maps the Boolean
description resulting from decoherence onto classical statistical mechanics.
\end{enumerate}

This shows that, strictly speaking, the classical limit of quantum mechanics
is not classical mechanics but classical statistical mechanics. This point
deserves some further remarks.

In the classical distribution $\rho _{c}(\phi )$ resulting from the
classical limit, the ensemble of trajectories is weighted by the
non-negative function $\rho (\omega ,p)$: it is precisely the fact that $%
\rho (\omega ,p)$ is non-negatively defined what permits it to be
interpreted as a probability function. But the formal agreement between $%
\rho _{c}(\phi )$ and a density distribution in standard classical
statistical mechanics does not mean that both have the same physical
meaning. In fact, in classical statistical mechanics probabilities are
conceived as a sort of measure of our ignorance about the real deterministic
classical trajectory. On the contrary, since the $\rho (\omega ,p)$ are the
diagonal components of the original quantum state $\rho $, they represent
quantum probabilities which, as many no-go theorems show, are irreducible.
Of course, this fact does not mean that a particular classical trajectory
cannot be chosen. Let us suppose that we prepare the system at $t=0$ in an
initial condition $\rho $ such that its singular part $\rho _{S}$ is an
eigenstate $(\omega ,p|$. In this case, as a consequence of decoherence and
macroscopicity, we will obtain the particular trajectory defined by the
constants of motion $H(\phi )=\omega $ and $P(\phi )=p$ with certainty. This
shows that, although $\rho (\omega ,p)$ in the classical distribution $\rho
_{c}(\phi )$ represents quantum irreducible probabilities, a particular
classical trajectory can always be selected by means of the proper
preparation of the quantum initial conditions.

Finally, it is worth stressing the emergent nature of classicality as
explained by the present approach. As we have seen, the off-diagonal terms
of the quantum state $\rho (t)$ never vanish through the unitary quantum
evolution. Strictly speaking, what self-induced approach shows is that, in
the infinite time limit, for any $A\in {\cal A}\,$, $\langle A\rangle _{\rho
(t)}$ can be computed {\it as if} the system were in the diagonal state ${%
\rho }_{*}$. In fact, the limit for $t\rightarrow \infty $ of $\langle
A\rangle _{\rho (t)}$ could also be computed in the Heisenberg picture as
the limit for $t\rightarrow \infty $ of $\langle A(t)\rangle _{\rho }$; in
this case we would obtain a diagonal operator $A_{*}$. This fact clearly
shows that the fundamental magnitude in the explanation of decoherence is $%
\langle A\rangle _{\rho (t)}=$ $\langle A(t)\rangle _{\rho }$ and not ${\rho 
}_{*}$ nor $A_{*}$. In other words, decoherence should be conceived as a
coarse-grained process that describes the evolution of the state $\rho (t)$
from the observational viewpoint given by the observable $A$\footnote{$%
\langle A\rangle _{\rho (t)}=(\rho |A)$ can be thought as a generalized
''projection'' of the state $\rho $. In fact, we can define a projector $\Pi 
$ belonging to ${\cal A}\otimes {\cal A}^{*}$ as $\Pi =A\rho _{A}$, where $%
\rho _{A}$ satisfies $(\rho _{A}|A)=1$ (this condition guarantees that $\Pi
^{2}=\Pi $). In this case, $\rho _{rel}=(\rho |A)\rho _{A}$, where $\rho
_{rel}$ is the projected part of $\rho $ relevant for decoherence. Since
coarse-graining amounts to a projection (see \cite{Mackey}), $\langle
A\rangle _{\rho (t)}$ can be conceived as a coarse-grained magnitude.}. As a
consequence, classicality is an emergent property that arises in a
coarse-grained level of description\footnote{%
An interesting discussion about emergence and reductionistic relations
between the various levels of the quantum mechanical descriptions can be
found in papers \cite{Primas}, \cite{Kronz}.}. The classical limit shows
that, from the point of view given by $A$, as the result of decoherence and
macroscopicity the quantum system behaves {\it as if} it were a classical
statistical system. This means that our measurements of the mean value of
any relevant observable $A$ on the quantum system will give the same results
as those we would obtain on a classical statistical system described as an
ensemble of classical trajectories weighted by their corresponding
probabilities. The distinction between the fundamental and the
coarse-grained levels of description permits us to understand how the
Boolean and deterministic classical world objectively arises from an
underlying non-Boolean and indeterministic quantum level.

\section{Conclusion}

Einstein was right when he considered the idea that the limit $\hbar
\rightarrow 0$ is the right classical limit as an oversimplification. On the
basis of the assumption that the problem of the classical limit of quantum
mechanics amounts to the question of how the classical world arises from an
underlying quantum reality, our account of the problem involves two
elements. The first one is self-induced decoherence, conceived as a process
that depends on the own dynamics of a closed quantum system governed by a
Hamiltonian with continuous spectrum; the study of decoherence was addressed
by means of formal tools derived from the van Hove formalism. The second
element is macroscopicity represented by the limit $\hbar \rightarrow 0$; we
have shown that, when the macroscopic limit is applied to the Wigner
transformation of the diagonal state resulting from decoherence, the
description of the quantum system becomes equivalent to the description of
an ensemble of classical trajectories on phase space weighted by their
corresponding probabilities. Furthermore, this approach to the classical
limit explains under what conditions the problems arising from the
traditional approach can be avoided. Finally, when these formal results are
considered in the light of a generalized concept of coarse-graining,
decoherence turns out to be a coarse-grained process that, in the infinite
time limit, leads to classicality when the system is macroscopic enough.
Since there is no subjective element involved in this process, from our
approach classicality is a property that objectively emerges from the
underlying quantum world.

\section{Acknowledgments}

We are very grateful to Olimpia Lombardi and Michael Zeitlin for their
valuable comments. This paper was partially financed by grants of CONICET,
the University of Buenos Aires, the Junta de Castilla y Le\'{o}n Project VA
085/02, the FEDER-Spanish Ministry of Science and Technology Projects DGI
BMF 2002-0200 and DGI BMF2002-3773.

\appendix

\section{The quantum evolution in classical terms}

The quantum evolution of the system can be translated into the classical
language via the Wigner transformation. The phase space analogue of the
Liouville-von Neumann equation for $\hbar \rightarrow 0$, eq.(\ref{4.10}),
is:

\begin{equation}
\partial _{t}\rho (t)=\{H,\rho \}_{mb}  \label{A1-1}
\end{equation}
Let us now compute $\rho (\phi ,t)=symb{\rho }(t)$, which is equal, up to
the order $\hbar ^{2}$ (hence, it is in fact the limit when $\hbar
\rightarrow 0$), to: 
\begin{equation}
\rho (\phi ,t)=\rho \left( H(\phi ),P(\phi )\right) +\int_{0}^{\infty
}\int_{0}^{\infty }\int_{p}\int_{p^{\prime }}\rho (\omega ,\omega ^{\prime
},p,p^{\prime })\,e^{i(\omega -\omega ^{\prime })t/\hbar }\,symb(\omega
,\omega ^{\prime },p,p^{\prime }|\,d\omega \,d\omega ^{\prime
}\,dp\,dp^{\prime }  \label{A1-2}
\end{equation}
Here we have used the Liouville equation for the regular part, and we have
kept the singular part unchanged since we know that it is time invariant. As
in the quantum case, we can call the first term the ``invariant part'' and
the second term the ``fluctuating part'' of $\rho (\phi ,t)$ \cite{Pra}, 
\cite{Rolo}. On this basis we can also compute: 
\begin{eqnarray}
(\rho (\phi ,t)|A(\phi )) &=&\int_{H}\int_{P}\,\rho (H(\phi ),P(\phi
))\,A(H(\phi ),P(\phi ))\,dH\,dP+  \nonumber \\
&&\int_{0}^{\infty }\int_{0}^{\infty }\int_{p}\int_{p^{\prime }}\rho (\omega
,\omega ^{\prime },p,p^{\prime })\,e^{i(\omega -\omega ^{\prime })t/\hbar
}\,A(\omega ,\omega ^{\prime },p,p^{\prime })\,d\omega \,d\omega ^{\prime
}\,dp\,dp^{\prime }  \label{A1-3}
\end{eqnarray}
where, again, we call the first term ``invariant part'' and the second term
``fluctuating part''. Now, if the product $\rho (\omega ,\omega ^{\prime
},p,p^{\prime })A(\omega ^{\prime },\omega ,p^{\prime },p)$ is integrable,
we can use the Riemann-Lebesgue theorem to conclude that: 
\begin{equation}
\lim_{t\rightarrow \infty }(\rho (\phi ,t)|A(\phi ))=\int_{H}\int_{P}\,\rho
(H(\phi ),P(\phi ))\,\,A(H(\phi ),P(\phi ))\,dH\,dP=(\rho _{*}(\phi )|A(\phi
))  \label{A1-4}
\end{equation}
for any $A(\phi )\in {\cal L}$ and for any $\rho (\phi )\in {\cal L}^{*}$
with the right properties. Therefore: 
\begin{equation}
\rho _{*}(\phi )=\int_{H}\int_{P}\rho (H(\phi ),P(\phi
))\,dH\,dP\,=\int_{0}^{\infty }\int_{p}\rho (\omega ,p)\,\delta (H(\phi
)-\omega )\,\delta (P(\phi )-p)\,d\omega \,dp  \label{A1-5}
\end{equation}
This last equation is precisely the Wigner transformation of $\rho _{*}$ for 
$\hbar \rightarrow 0$, as obtained in eq.(\ref{5.16}). In $\rho _{*}(\phi )$
the non-diagonal terms have disappeared and only the diagonal (singular)
terms remain. Thus, we have found the weak limit:

\begin{equation}
\qquad w-\lim_{t\rightarrow \infty }\rho (\phi ,t)=\rho _{*}(\phi )
\label{A1-6}
\end{equation}
which express the result of decoherence.

\section{Positivity of the Wigner function of $\rho (\omega )$}

Here we will prove that the Wigner function of a state represented by a
density operator of the form $\rho (\omega )=\rho (\omega ,\omega )$ is
positively defined a.e. in the limit $\hbar \rightarrow 0$. This proof is a
reformulation of an argument due to Narcovich \cite{Kastler}, \cite
{Narcovich}.

Let us call $\rho _{\hbar }(q,p)$ the Wigner function for the density
operator $\rho $ in terms of $\hbar $. We will call:

\begin{equation}
\lim_{\hbar \rightarrow 0}\rho _{\hbar }(q,p)=G(q,p)\,  \label{A2-1}
\end{equation}
We will prove that $G(q,p)$ is non-negative (a.e.):

\begin{equation}
G(q,p)\ge 0  \label{A2-2}
\end{equation}
Let us call:

\begin{equation}
a=\left( 
\begin{array}{c}
q \\[2ex] 
p
\end{array}
\right) \,,\qquad z=\left( 
\begin{array}{c}
q^{\prime } \\[2ex] 
p^{\prime }
\end{array}
\right) \,,\qquad J=\left( 
\begin{array}{cc}
0 & 1 \\[2ex] 
-1 & 0
\end{array}
\right)  \label{A2-3}
\end{equation}
and

\begin{equation}
\sigma (a,z)=aJz=(q,p)\,\left( 
\begin{array}{cc}
0 & 1 \\[2ex] 
-1 & 0
\end{array}
\right) \,\left( 
\begin{array}{c}
q^{\prime } \\[2ex] 
p^{\prime }
\end{array}
\right) =qp^{\prime }-pq^{\prime }  \label{A2-4}
\end{equation}
Now we consider the inverse symplectic Fourier transformation of $\rho
_{\hbar }(q,p)=\rho _{\hbar }(\phi )$ given by:

\begin{equation}
f_{\hbar }(z)=\int \rho _{\hbar }(\phi )\,e^{i\sigma (\phi ,z)}\,dz
\label{A2-5}
\end{equation}
where $dz=d{\bf q}\,d{\bf p}$. This transformation (i.e. the function $%
f_{\hbar }(z)$) is of $\hbar $-positive type (see \cite{Kastler}, \cite
{Narcovich}), which means that:

\begin{equation}
\sum_{j,k=1}^{m}f_{\hbar }(a_{j}-a_{k})\,e^{i(\hbar /2)\,\sigma
(a_{k},a_{j})}\,\lambda _{j}^{*}\lambda _{k}\ge 0  \label{A2-6}
\end{equation}
where:

1.- $a_{1},a_{2},\dots ,a_{m}$ are arbitrary points of the phase space:

\begin{equation}
a_{k}=\left( 
\begin{array}{c}
q_{k} \\[2ex] 
p_{k}
\end{array}
\right)  \label{A2-7}
\end{equation}

2.- $\lambda _{1},\lambda _{2},\dots ,\lambda _{m}$ are arbitrary numbers.

3.- $m=1,2,\dots $ is an arbitrary finite positive integer.

Let us now consider:

\begin{equation}
g(z)=\lim_{\hbar \rightarrow 0}f_{\hbar }(z)  \label{A2-8}
\end{equation}
and see whether $g(z)$ is positive in the sense of Bochner. If we make the
limit $\hbar \rightarrow 0$ in eq.(\ref{A2-5}), we obtain:

\begin{equation}
\sum_{j,k=1}^{m}g(a_{j}-a_{k})\,\lambda _{j}^{*}\lambda _{k}\ge 0
\label{A2-9}
\end{equation}
where the notation has been defined in 1, 2, 3 right above. Thus, the
function $g(z)$ is positive in the sense of Bochner. Therefore, the Fourier
transform $\varphi (q,p)$ is also positive (a.e.):

\begin{equation}
\varphi (q,p)=\int g(z)\,e^{-iza}\,dz\ge 0  \label{A2-10}
\end{equation}
This property is inherited by the symplectic Fourier transform of $g(z)$,
that we have called $G(q,p)$:

\begin{equation}
G(q,p)=\varphi (p,-q)  \label{A2-11}
\end{equation}
Then, from eq.(\ref{A2-8}) we have:

\begin{eqnarray}
\lim_{\hbar \mapsto 0}\rho _{\hbar }(q,p) &=&\int \left[ \lim_{\hbar \mapsto
0}f_{\hbar }(z)\right] \,e^{-i\sigma (a,z)}\,dz  \nonumber \\[0.03in]
&=&\int g(z)\,e^{-i\sigma (a,z)}\,dz=G(q,p)\ge 0\qquad (a.e)  \label{A2-12}
\end{eqnarray}
where $\rho _{\hbar }(q,p)$ is an arbitrary regular Wigner function, q.e.d.
Nevertheless, it is worth stressing that the non-negativeness obtained
through decoherence is stronger than this result: $\rho _{c}(\phi )$ is
non-negatively defined on the whole phase space, and not only a.e.

\vfill\eject

\end{document}